\documentclass[twocolumn,showpacs,amsmath,amssymb,prc]{revtex4-1}

\usepackage[dvipdfmx]{graphicx}
\usepackage{dcolumn}
\usepackage{bm}
\usepackage{CJK}
\usepackage{braket}

\begin{document}


\title{
$^{16}$O + $^{16}$O molecular structures of positive- and negative-parity superdeformed bands in $^{34}$S
}

\begin{CJK}{UTF8}{min}
 \author{Yasutaka Taniguchi (谷口\ 億宇)}
 \affiliation{Center for Computational Sciences, University of Tsukuba, 1-1-1 Tennohdai, Tsukuba, Ibaraki 305-8577, Japan.}
  \affiliation{Department of Medical and General Sciences, Nihon Institute of Medical Science, 1276 Shimogawara, Moroyama-machi, Iruma-gun, Saitama 350-0435, Japan.}

 \date{\today}

 \begin{abstract}
  The structures of excited states in $^{34}$S are investigated using the antisymmetrized molecular dynamics and generator coordinate method (GCM).
  The GCM basis wave functions are calculated via energy variation with a constraint on the quadrupole deformation parameter $\beta$.
  By applying the GCM after parity and angular momentum projections, the coexistence of two positive- and one negative-parity superdeformed (SD) bands are predicted, and low-lying states and other deformed bands are obtained.
  The SD bands have structures of $^{16}$O + $^{16}$O + two valence neutrons in molecular orbitals around the two $^{16}$O cores in a cluster picture.
  The configurations of the two valence neutrons are $\delta^2$ and $\pi^2$ for the positive-parity SD bands and $\pi^1\delta^1$ for the negative-parity SD band.
  The structural changes of the yrast states are also discussed.
 \end{abstract}

 \pacs{ }
 \maketitle
\end{CJK}

\section{Introduction}

Dynamic structural changes under excitation are significant properties of nuclei.
Superdeformation and clustering are typical changes.
With the development of techniques for $\gamma$ spectroscopy experiments, superdeformed (SD) bands have been observed in the $A\sim 40$ region, for example, in $^{36,38,40}$Ar\cite{PhysRevC.63.061301,PhysRevC.65.034305,Ideguchi201018}, $^{35}$Cl\cite{PhysRevC.88.034303}, $^{40,42}$Ca\cite{PhysRevLett.87.222501,PhysRevC.67.041303,springerlink:10.1007/BF01412107,NSR2003LA04}, and $^{44}$Ti\cite{PhysRevC.61.064314}, and microscopic theoretical studies have shown that these bands have multiparticle-multihole (mp-mh) excited structures.
Clustering is a typical structure in the light mass region, for example, in $^{8}$Be, $^{12}$C, $^{16}$O, and $^{20}$Ne\cite{PTPS.52.89,PTPS.68.29}.
In the $A\sim 30\mbox{-}40$ region, cluster correlations in highly deformed states have also been discussed\cite{PhysRevC.66.021301,PhysRevC.66.021301,PhysRevC.69.051304,Kimura200658,taniguchi:044317,PhysRevC.80.044316}.
To clarify the dynamic structural changes of nuclei, it is necessary to study the nuclear structure in terms of both deformation and clustering.
However, those studies have been insufficient.

S isotopes $(Z = 16)$ are suitable nuclei for studying deformation and clustering caused by excitation.
S isotopes are expected to be favorable for the formation of SD bands because $Z = 16$ is considered a magic number of superdeformation.
The existence of SD bands in S isotopes has been discussed in the frameworks of a mean-field model\cite{Inakura200352}, a cluster model\cite{PhysRevC.66.021301}, and antisymmetrized molecular dynamics (AMD)\cite{PhysRevC.69.051304}.
In terms of clustering, S isotopes are key nuclei in the $sd$ shell.
S isotopes are analogs of Be isotopes because those isotopes can form systems consisting of two doubly closed shell nuclei ($^{16}$O and $\alpha$ for S and Be isotopes, respectively) and valence neutrons.
In Be isotopes, structures consisting of $\alpha$ + $\alpha$ + valence neutrons in molecular orbitals are thought to develop in low-lying states, with the valence neutrons in molecular orbitals around two $\alpha$ cores\cite{Seya01011981,PhysRevC.60.064304,PhysRevC.61.044306,Ito200443,Ito2006293,PhysRevLett.100.182502,PhysRevC.85.014302,Kanada-En'yo01012012}.
The SD states in $^{32}$S are predicted to contain many $^{16}$O + $^{16}$O cluster structure components\cite{PhysRevC.66.021301,PhysRevC.69.051304}.
They suggest the existence of SD states that have $^{16}$O + $^{16}$O + valence neutrons in the molecular orbital structure in S isotopes.

By a $\gamma$ spectroscopy experiment, the structures in $^{34}$S are investigated up to the $J^\pi = 10^+$ and $(9^-)$ states for positive- and negative-parity states, respectively, mainly for the yrast states\cite{PhysRevC.71.014316}.
The $B(\mathrm{E2})$ value of the transition $8^+_1$ (10.650 MeV) $\rightarrow$ $6^+$ (8.503 MeV) is $27\pm 15~B_{\mathrm{Wu}}(\mathrm{E2})$\cite{PhysRevC.71.014316}, which is large enough that the $J^\pi = 6^+$ and $8^+$ states can be interpreted as members of a rotational band, where $B_\mathrm{Wu}(\mathrm{E2})$ is the Weisskopf unit.
Analysis using a shell model shows that the yrast states for $J^\pi \geq 6^+$ are $2\hbar\omega$ excited states, whereas those for $J^\pi \leq 4^+$ have $0\hbar\omega$ configurations.
For the negative-parity states, a shell model shows that the yrast states have $1\hbar\omega$ configurations up to $J^\pi = 9^-$.
In contrast to the yrast states, the structures of the non-yrast states with mp-mh configurations have never been clarified.

$^{34}$S is an analog of $^{10}$Be because both isotopes can form a system consisting of two doubly closed shell nuclei and two valence neutrons.
In low-lying states in $^{10}$Be, structures consisting of $\alpha$ + $\alpha$ + two valence neutrons in molecular orbitals are thought to develop\cite{Seya01011981,PhysRevC.60.064304,PhysRevC.61.044306,Ito2006293}.
The molecular orbitals around the $\alpha$ + $\alpha$ are formed by linear combination of $0p_{3/2}$ orbits around the two $\alpha$ cores.
The configurations of the valence neutrons are considered to be $\pi^2$, $\sigma^2$, and $\pi^1 \sigma^1$ for the $J^\pi = 0_1^+$, $0_2^+$, and $1_2^-$ states, respectively\cite{PhysRevC.60.064304}.
In $^{34}$S, the candidate configurations of the molecular orbitals around $^{16}$O + $^{16}$O are the $\delta$, $\pi$, and $\sigma$ orbitals, which are formed by linear combinations of $0d_{5/3}$ orbits around the two $^{16}$O cores.
The structures of the low-lying states in $^{10}$Be suggest the coexistence of positive- and negative-parity SD states in $^{34}$S with $^{16}$O + $^{16}$O + valence neutrons in molecular orbitals.
Superdeformation in S isotopes has been discussed systematically using mean-field calculations\cite{Inakura200352}, but the detailed structures have never been discussed in $^{34}$S.
Superdeformation and clustering in $^{34}$S are open problems.

This paper aims to clarify the structures of SD states in $^{34}$S using AMD and the generator coordinate method (GCM).
The coexistence of positive- and negative-parity SD bands and their structures are discussed, focusing on $^{16}$O + $^{16}$O + valence neutrons in molecular orbitals around the two $^{16}$O cores.
The structural changes in the yrast states are also discussed.

This paper is organized as follows.
In Sec.~\ref{framework}, the framework of this study is explained briefly.
In Sec.~\ref{results}, the numerical results are presented.
In Sec.~\ref{discussions}, the structures of the SD and yrast states are discussed.
Finally, conclusions are given in Sec.~\ref{conclusions}.

\section{Framework}
\label{framework}

In this section, the framework of the study is explained briefly.
The details of the framework are provided in Refs.~\onlinecite{PTP.93.115,PhysRevC.69.044319,PTP.112.475}.

\subsection{Wave function}

The wave functions in low-lying states are obtained using parity projection
and angular momentum projection (AMP) and the GCM with deformed-basis AMD wave functions. A deformed-basis AMD wave function
$\ket{\mathrm{\Phi}}$ is a Slater determinant of Gaussian wave
packets that can deform triaxially such that
\begin{eqnarray}
 &\ket{\mathrm{\Phi}} = \hat{\cal A}\ket{\varphi_1, \varphi_2, \cdots, \varphi_A},&\\
 &\ket{\varphi_i} = \ket{\phi_i} \otimes \ket{\chi_i} \otimes \ket{\tau_i},& \\
 &\langle \mathbf{r} | \phi_i \rangle = \pi^{-3/4} (\det {\mathsf K})^{1/2} \exp\left[ - \frac{1}{2} ({\mathsf K} \mathbf{r} - \mathbf{Z}_i)^2\right],& \\
 &\ket{\chi_i} = \chi^\uparrow_i \ket{\uparrow} + \chi^\downarrow_i \ket{\downarrow},& \\
 &\ket{\tau_i} = \ket{\pi}\ \mathrm{or}\ \ket{\nu},&
\end{eqnarray}
where $\hat{\cal A}$ denotes the antisymmetrization operator, and
$\ket{\varphi_i}$ denotes a single-particle wave function. Further,
$\ket{\phi_i}$, $\ket{\chi_i}$, and $\ket{\tau_i}$ denote the
spatial, spin, and isospin components, respectively, of each
single-particle wave function $\ket{\varphi_i}$. The real $3 \times
3$ matrix $\mathsf{K}$ denotes the width of the Gaussian
single-particle wave functions that can deform triaxially and is
common to all nucleons. $\mathbf{Z}_i = (Z_{ix}, Z_{iy}, Z_{iz})$ are complex parameters denoting the centroid of
each single-particle wave function in phase space. The complex
parameters $\chi_i^\uparrow$ and $\chi_i^\downarrow$ denote the spin
directions. Axial symmetry is not assumed.

\subsection{Energy variation}

The basis wave functions of the GCM are obtained by energy
variation with a constraint potential $V_\mathrm{cnst}$ after projection onto eigenstates of parity,
\begin{eqnarray}
 &\delta \left( \frac{\braket{\Phi^\pi | \hat{H} | \Phi^\pi}}{\braket{\Phi^\pi | \Phi^\pi}} + V_{\mathrm{cnst}} \right) = 0,&\\
 &\ket{\Phi^\pi} = \frac{1 + \pi \hat{P}_r}{2} \ket{\Phi},&
\end{eqnarray}
where $\hat{H}$ and $\hat{P}_r$ denote the Hamiltonian and parity operator, respectively.
The variational parameters are
$\mathsf{K}$, $\mathbf{Z}_i$, and $\chi_i^{\uparrow,\downarrow}$ ($i= 1, ..., A$). The isospin component of each single-particle wave
function is fixed as a proton ($\pi$) or a neutron ($\nu$). The
Gogny D1S force is used as the effective interaction. 

To obtain the deformed wave functions, the constraint potential $V_{\mathrm{cnst}}$ for the matter quadrupole deformation parameter $\beta$ of the total system is used.

\subsection{Generator coordinate method}

  The optimized wave functions are superposed after parity projection and AMP by employing the quadrupole deformation parameter $\beta$,
 \begin{equation}
    \Ket{\Phi^{J\pi}_M} = \sum_i f_i \hat{P}_{MK_i}^{J\pi}  \Ket{\Phi(\beta_i)}, \label{eq:HW} 
 \end{equation} 
where $\hat{P}_{MK}^{J^\pi}$ is the parity and total angular momentum
  projection operator, and $\Ket{\Phi(\beta_i)}$ are optimized wave functions with a constraint for $\beta = \beta_i$.
  The integrals over the three Euler angles in the total angular momentum projection operator $\hat{P}_{MK}^J$ are evaluated by numerical integration.  
  The numbers of sampling points in the numerical integration are 23, 27, and 23 for $\alpha$, $\beta$, and $\gamma$, respectively. 
  Here the body-fixed $x$, $y$, and $z$ axes are chosen as $\langle x^2 \rangle \leq \langle y^2 \rangle \leq \langle z^2 \rangle$ for the $\gamma < 30^\circ$ wave functions and  $\langle x^2 \rangle \geq \langle y^2 \rangle \geq \langle z^2 \rangle$ for the $\gamma > 30^\circ$ ones. 
  The coefficients $f_i$ are
  determined by the Hill--Wheeler equation,  
  \begin{equation}
    \delta \left( \Braket{\Phi^{J\pi}_M | \hat{H} | \Phi^{J\pi}_M} - \epsilon \Braket{\Phi^{J\pi}_M | \Phi^{J\pi}_M}\right) = 0. 
  \end{equation}
  Then we obtain the energy spectra and the corresponding wave functions, which are expressed by the superposition of the optimum wave functions, $\{ \ket{\Phi(\beta_i)} \}$.

\section{Results}
\label{results}

\subsection{Energy variation}

\subsubsection{Energy curves}

\begin{figure}[tbp]
 \begin{center}
  \includegraphics[width=0.5\textwidth]{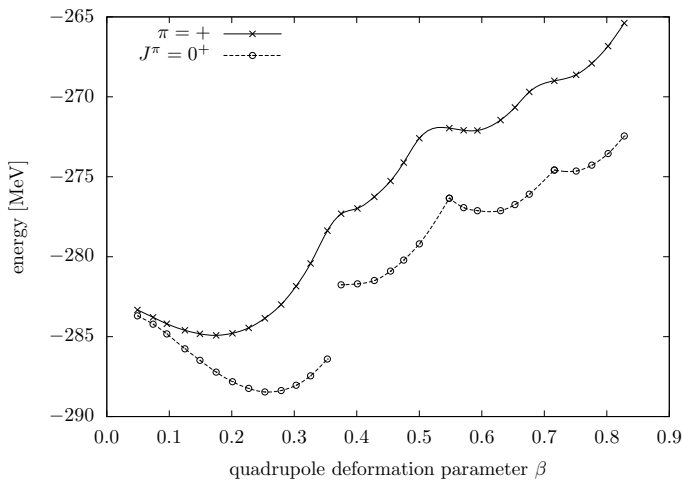}\\
  \caption{
  Energy surfaces as functions of quadrupole deformation parameter $\beta$ for positive-parity states.
  Solid and dashed lines show energies projected onto positive-parity and $J^\pi = 0^+$ states, respectively.
  }
  \label{beta_energy+}
 \end{center}
\end{figure}

\begin{figure}[tbp]
 \begin{center}
  \includegraphics[width=0.5\textwidth]{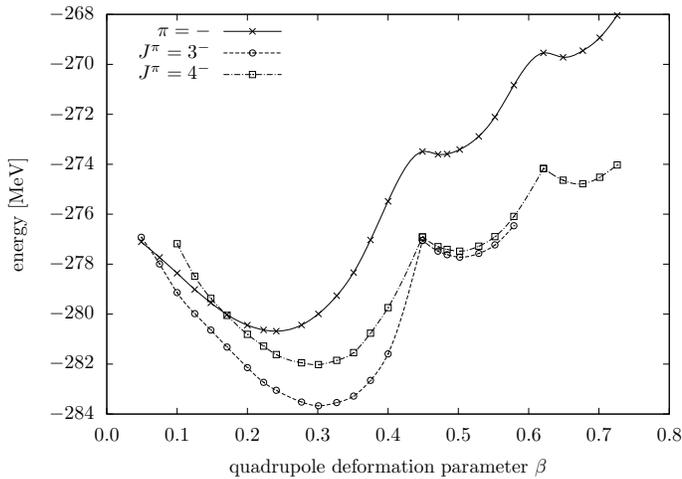}\\
  \caption{
  Energy surfaces as functions of quadrupole deformation parameter $\beta$ for negative-parity states.
  Solid, dashed, and dot-dashed lines show energies projected onto positive-parity, $J^\pi = 3^-$, and $J^\pi = 4^-$ states, respectively.
  }
  \label{beta_energy-}
 \end{center}
\end{figure}

Figures~\ref{beta_energy+} and \ref{beta_energy-} show the energy surfaces as functions of the quadrupole deformation parameter $\beta$ for the positive- and negative-parity states, respectively, obtained by energy variation with a constraint on $\beta$ after parity projection.
The energies projected onto the $J^\pi = 0^+$, $3^-$, and $4^-$ states are also shown.

In the positive-parity energy surface (Fig.~\ref{beta_energy+}), three excited local minima or shoulders exist around $\beta = 0.4$, 0.6, and 0.8 as well as the minimum at $\beta = 0.25$, which suggest the existence of three excited deformed bands in the positive-parity states.
After AMP onto the $J^\pi = 0^+$ states, more deformed states gain more binding energy, and the $\beta$ values of the local minima become larger. 
%
In the negative-parity energy surface (Fig.~\ref{beta_energy-}), three local minima exist around $\beta = 0.2$, 0.5, and 0.7.
In the slightly deformed region, $\beta < 0.6$, the $J^\pi = 3^-$ components have the lowest energies after AMP.
Highly deformed wave functions, $\beta > 0.6$, have few $J^\pi = 3^-$ components, and the $J^\pi = 4^-$ components have the lowest energies.

\subsubsection{Particle-hole configuration of deformed states}

\begin{figure}[tbp]
 \begin{center}
  \includegraphics[width=0.5\textwidth]{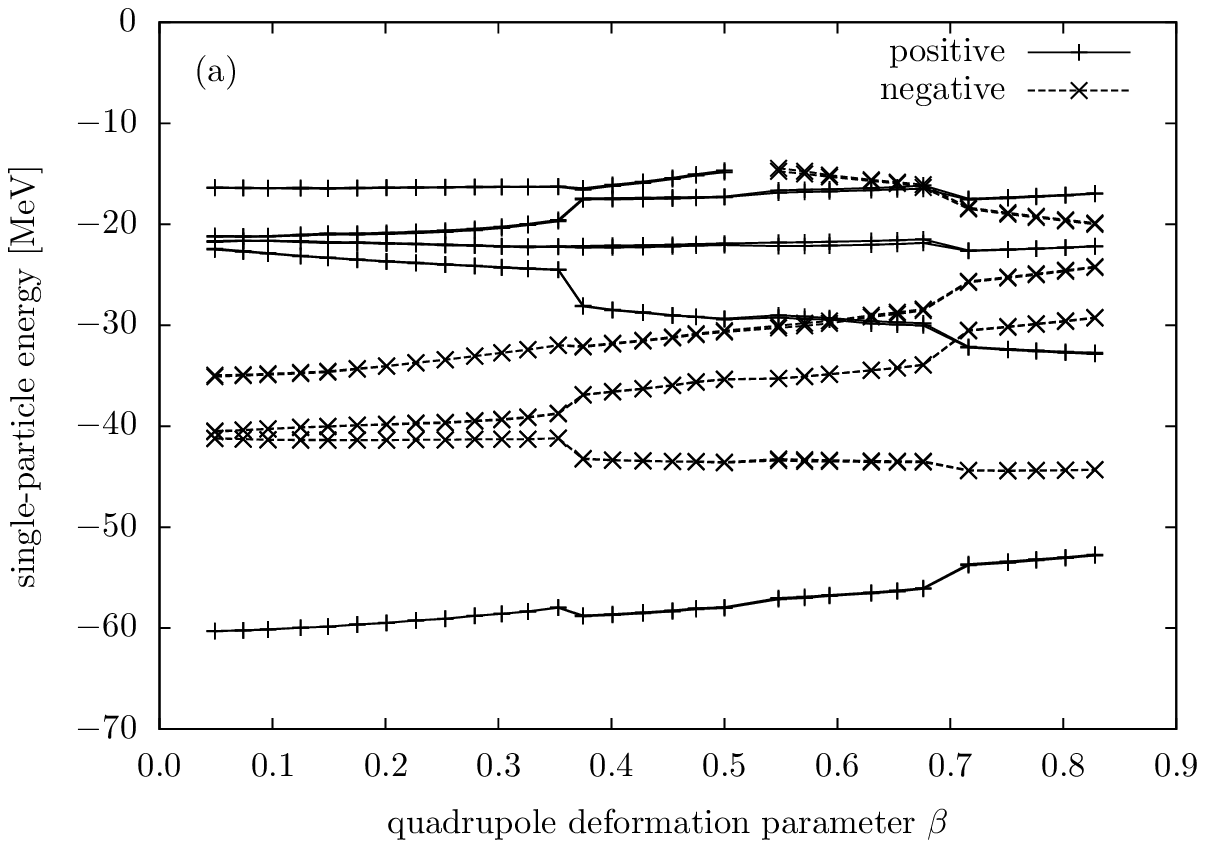}\\
  \includegraphics[width=0.5\textwidth]{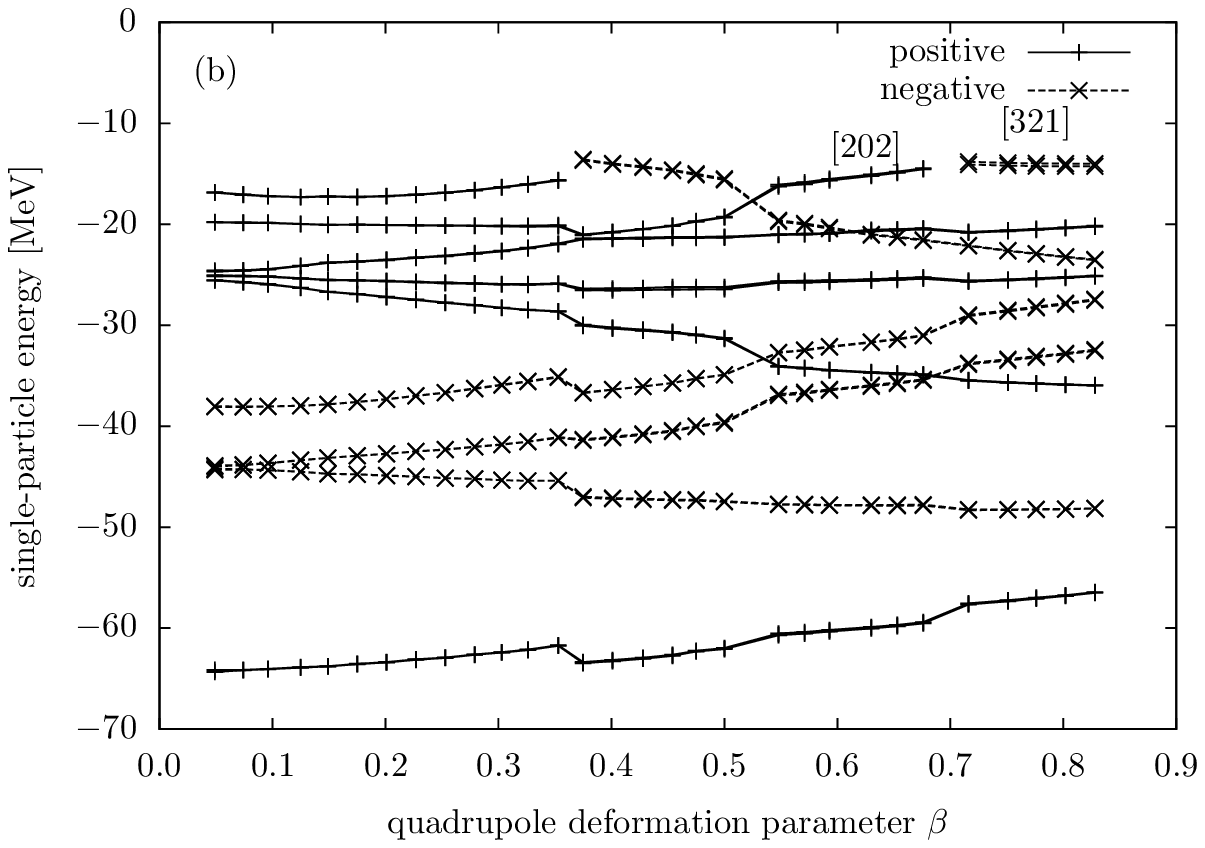}\\
  \caption{
  Single-particle energies of (a) protons and (b) neutrons as functions of quadrupole deformation parameter $\beta$ for positive-parity states.
  Solid and dashed lines show positive- and negative-parity orbits, respectively.
  Numbers in  brackets show  the Nilsson quanta for the two highest orbits of neutrons (see text).
  }
  \label{spo_positive}
 \end{center}
\end{figure}

\begin{figure}[tbp]
 \begin{center}
  \includegraphics[width=0.5\textwidth]{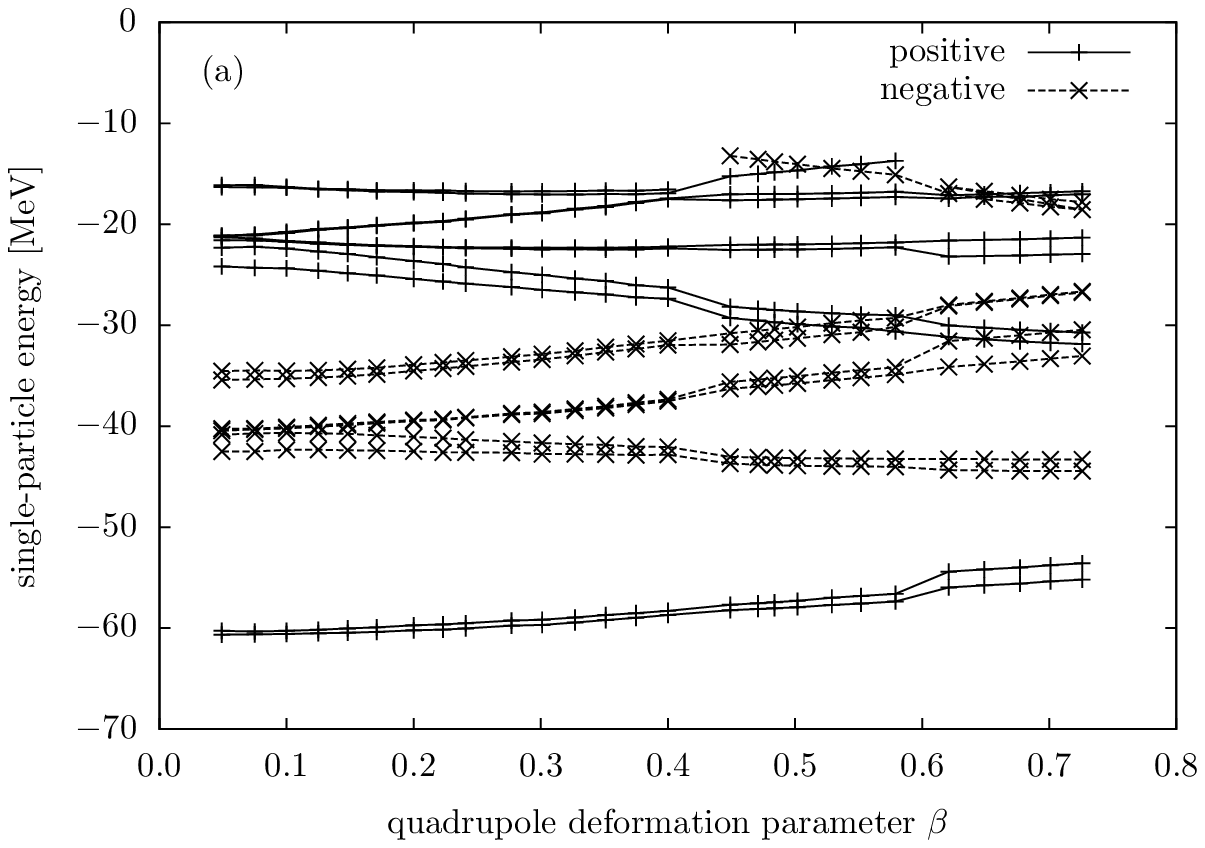}\\
  \includegraphics[width=0.5\textwidth]{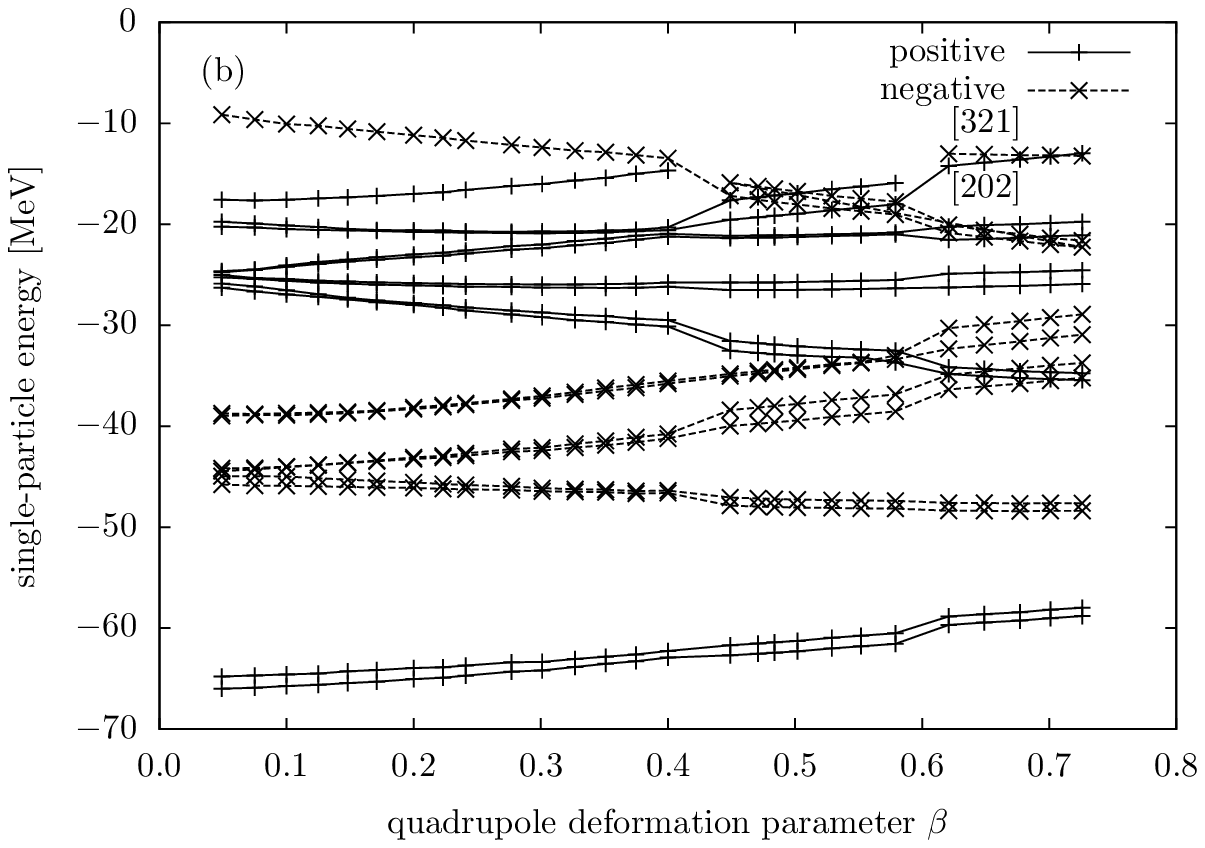}\\
  \caption{Same as Fig.~\ref{spo_positive} but for negative-parity states.}
  \label{spo_negative}
 \end{center}
\end{figure}

Figures~\ref{spo_positive} and \ref{spo_negative} show the single-particle energies as functions of the quadrupole deformation $\beta$ for positive- and negative-parity states, respectively.
The quanta $[N n_z \Omega]$ in the Nilsson picture are also shown for the two highest orbits of neutrons in the highly deformed region.
The particle-hole configurations change dramatically depending on its deformation.

In the positive-parity states (Fig.~\ref{spo_positive}), two orbits are degenerate because of time-reversal symmetry.
The particle-hole configurations change at $\beta = 0.36$, 0.52, and 0.7.
In the slightly deformed region ($\beta < 0.36$), the wave functions have the lowest allowed configurations.
At $\beta = 0.36$, the neutron orbits originating in the $sd$ and $pf$ shells cross, and two neutrons move from orbits originating in the $sd$ shell to orbits originating in the $pf$ shell, which indicates $2\hbar\omega$ excitation in a spherical shell model picture.
At $\beta = 0.52$ and $0.7$, two protons and neutrons move from orbits originating in the $sd$ shell to orbits originating in the $pf$ shell.
The particle-hole configurations of all the positive-parity states are $(n_\pi, n_\nu) = (0, 0), (0, 2), (2, 2)$, and $(2, 4)$ for $\beta < 0.36$, $0.36 < \beta < 0.52$, $0.52 < \beta < 0.7$, and $\beta > 0.7$, respectively, where $n_\pi$ and $n_\nu$ are the numbers of protons and neutrons, respectively, in the single-particle orbit originating from the $pf$ shell.

\begin{figure}[tbp]
 \begin{center}
  \begin{tabular}{cc}
   \includegraphics[width=0.225\textwidth]{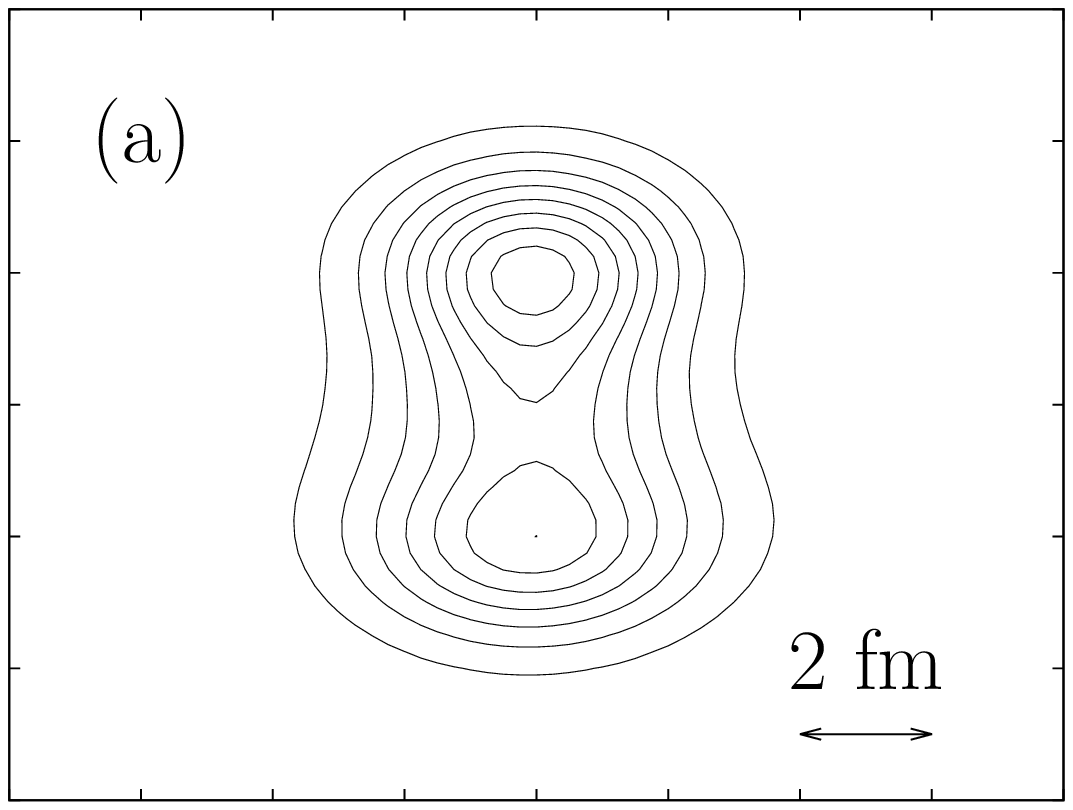} &
   \includegraphics[width=0.225\textwidth]{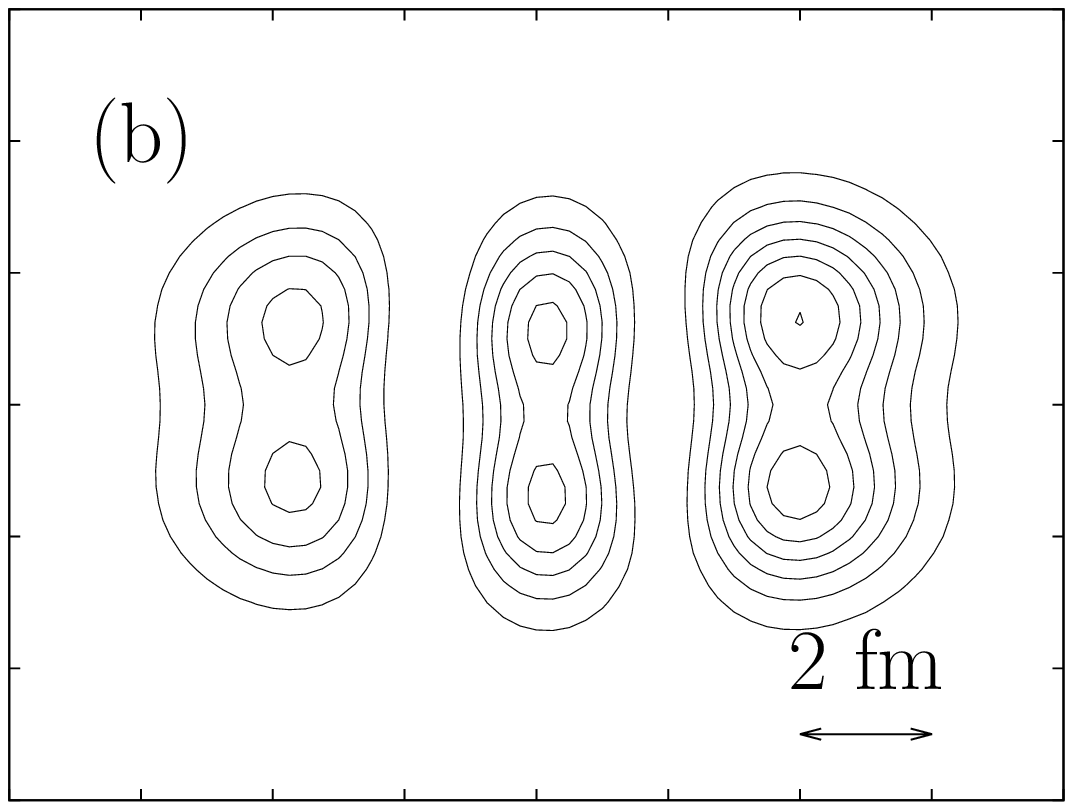} \\
   \includegraphics[width=0.225\textwidth]{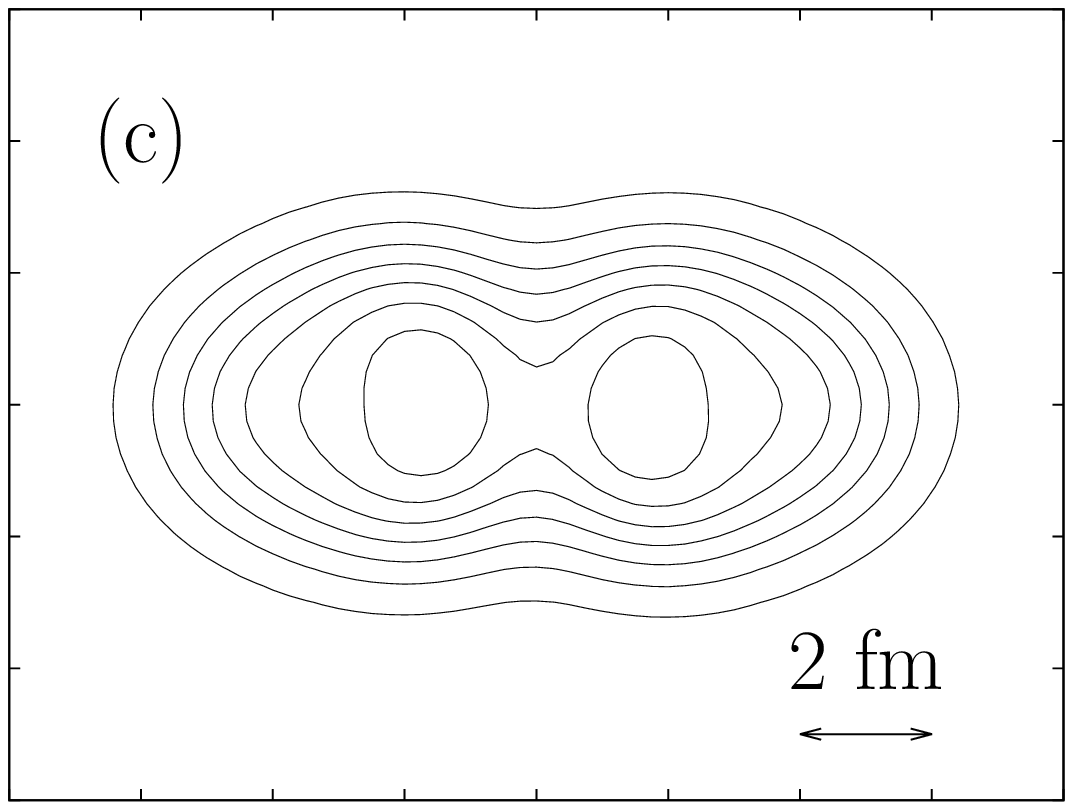}& \\
  \end{tabular}
  \caption{
  Density distributions of two highest single-particle orbits at (a) $\beta = 0.63$ and (b) $0.75$  and that of (c) proton component at $\beta = 0.75$ in positive-parity states.
  }
  \label{34S_density}
 \end{center}
\end{figure}

Figures~\ref{34S_density}(a) and (b) show the density distributions of the highest neutron orbits at $\beta = 0.63$ and 0.75, which are called $\psi^\mathrm{spo}_{0.63}$ and $\psi^\mathrm{spo}_{0.75}$, respectively.
The $\psi^\mathrm{spo}_{0.63}$ and $\psi^\mathrm{spo}_{0.75}$ orbits have no and two nodes in the $z$ direction (horizontal axis), respectively.
A calculation of the number of $l_z$ components,
\begin{equation}
 \Braket{\frac{1}{2\pi} \int_{-\pi}^{\pi} \mathrm{d}\phi~e^{-i (\hat{l}_z - l_z)\phi}},
\end{equation}
reveals that the $\psi^\mathrm{sop}_{0.63}$ and $\psi^\mathrm{spo}_{0.75}$ orbits contain dominantly $|l_z| = 2$ and 1 components, respectively, where the $z$ axis is chosen to be the long axis of the entire system.
The $\psi^\mathrm{spo}_{0.63}$ and $\psi^\mathrm{spo}_{0.75}$ orbits have no node on the radial coordinate in cylindrical coordinates
and therefore have [202] and [321] configurations, respectively, in the Nilsson picture. The Nilsson configurations of the two highest neutron orbits are $[202]^2$ and $[321]^2$ for $0.52 < \beta < 0.7$ and $\beta > 0.7$, respectively.
The single-particle energies of $[202]$ and $[321]$ are flat for the quadrupole deformation parameter $\beta$, and they resemble each other.

In the negative-parity states (Fig.~\ref{spo_negative}), the degeneracy is resolved by breaking of the time-reversal symmetry.
The particle-hole configurations change at $\beta = 0.4$ and 0.6.
At $\beta < 0.4$, the highest neutron orbit originates in the $pf$ shell and is a $1\hbar\omega$ excited configuration in a spherical shell model picture. 
The $sd$-shell-oriented and $pf$-shell-oriented orbits cross at $\beta = 0.4$ and 0.6 for protons and neutrons, respectively.
The particle-hole configurations of all the negative-parity states are $(n_\pi, n_\nu) = (0, 1), (1, 2)$, and $(2, 3)$ for $\beta < 0.4$, $0.4 < \beta < 0.6$, and $\beta > 0.6$, respectively. 
The Nilsson configurations of the two highest neutron orbits are $[202]^1$ and $[321]^1$ for $\beta > 0.6$.
The energies of the $[202]$ and $[321]$ orbits are flat for the quadrupole deformation parameter $\beta$, and they are almost the same.

The proton components of the $n_\pi = 2$ wave functions have similar density distributions for both positive- ($\beta > 0.5$) and negative-parity states ($\beta > 0.6$), and they have neck structures, as shown in Fig.~\ref{34S_density}(c). 
The particle-hole configurations and density distributions of the lower 16 neutrons are similar to those of protons in the highly deformed region.

\subsection{Level scheme}

\begin{figure}[tbp]
 \begin{center}
  \includegraphics[width=0.5\textwidth]{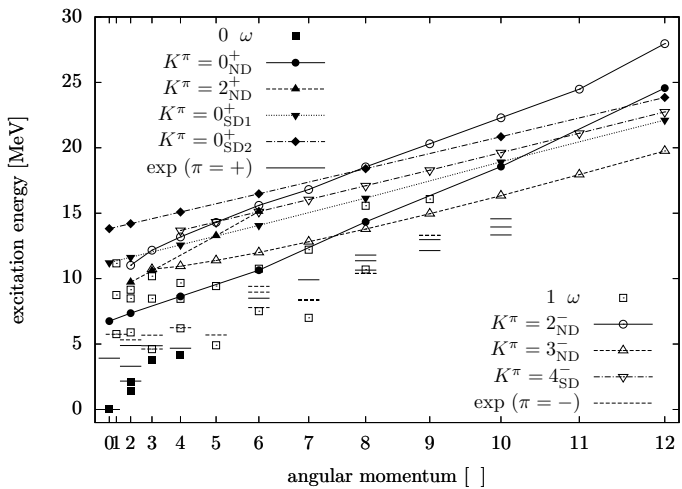}
  \caption{
  Excitation energies as functions of $J(J + 1)$, where $J$ is the total angular momentum.
  States with closed and open squares have dominant $0\hbar\omega$ and $1\hbar\omega$ configurations, respectively.
  Closed circles, triangles, inverted triangles, and diamonds show members of $K^\pi = 0_\mathrm{ND}^+$, $2_\mathrm{ND}^+$, $0_\mathrm{SD1}^+$, and $0_\mathrm{SD2}^+$ bands, respectively.
  Open circles, triangles, and inverted triangles show members of $2_\mathrm{ND}^-$, $3_\mathrm{ND}^-$, and $4_\mathrm{SD}^-$ bands, respectively.
  }
  \label{level_scheme}
 \end{center}
\end{figure}

Figure~\ref{level_scheme} shows the level scheme of the positive- and negative-parity states.
Various rotational bands, called $K^\pi = 0_\mathrm{ND}^+$, $0_\mathrm{SD1}^+$, and $0_\mathrm{SD2}^+$ for positive parity and $K^\pi = 2_\mathrm{ND}^-$, $4_\mathrm{SD}^-$, and $3_\mathrm{SD}^-$ for negative parity, are obtained, as well as low-lying $0\hbar\omega$ and $1\hbar\omega$ states.
The energies of the $0\hbar \omega$ and $1\hbar \omega$ states are consistent with experimental data.

In the positive-parity states, the yrast states have the $(n_\pi, n_\nu) = (0, 0)$ configuration with little deformation up to $J^\pi \leq 4^+$. 
The members of the $K^\pi = 0_\mathrm{ND}^+$ band, the dominant components of which are $(n_\pi, n_\nu) = (0, 2)$, become yrast states for $6^+ \leq J^\pi \leq 10^+$.
In high-spin states, $J^\pi \geq 12^+$, the members of the $K^\pi = 0_\mathrm{SD1}^+$ band are yrast states.
The energies of the $K^\pi = 0_\mathrm{SD2}^+$ band are a few MeV higher than those of the $K^\pi = 0_\mathrm{SD1}^+$ band.
The dominant components of the $K^\pi = 0_\mathrm{SD1}^+$ and $0_\mathrm{SD2}^+$ bands have $(n_\pi, n_\nu) = (2, 2)$ and $(2, 4)$ configurations, which are $4\hbar\omega$ and $6\hbar\omega$ excited configurations in a spherical shell model picture, respectively.

In the negative-parity states, the yrast states have $(n_\pi, n_\nu) = (0, 1)$ configurations for the $J^\pi \leq 8^-$ states.
For the $J^\pi \geq 9^-$ states, the members of a $K^\pi = 3^-$ band, called the $K^\pi = 3_\mathrm{ND}^-$ band, are yrast states.
Furthermore, $K^\pi = 2^-$ and $4^-$ bands exist, which are called the $K^\pi = 2^-_\mathrm{ND}$ and $4_\mathrm{SD}^-$ bands, respectively.
The dominant components of the $K^\pi = 3_\mathrm{ND}^-$ and $2_\mathrm{ND}^-$ bands have $(n_\pi, n_\nu) = (1, 2)$ configurations, and those of the $K^\pi = 4_\mathrm{SD}^-$ band have $(n_\pi, n_\nu) = (2, 3)$ components.
The $(n_\pi, n_\nu) = (1, 2)$ and $(2, 3)$ configurations are $3\hbar\omega$ and $5\hbar\omega$ configurations, respectively, in a spherical shell model picture.

The members of the $K^\pi = 0_\mathrm{SD1}^+$, $0_\mathrm{SD2}^+$, and $4_\mathrm{SD}^-$ bands have mp-mh configurations for both the proton and neutron components, and the values of the quadrupole deformation parameter $\beta$ of those dominant components are greater than 0.6.
Their energies are within a few MeV, although the particle-hole configurations of the neutron components differ; they are $n_\nu = 2$, 3, and 4 for the $K^\pi = 0_\mathrm{SD1}^+$, $4_\mathrm{SD}^-$, and $0_\mathrm{SD2}^+$ bands, respectively.
This shows the coexistence of two positive- and one negative-parity SD bands.
The energy of the $K^\pi = 4_\mathrm{SD}^-$ band is intermediate between those of the $K^\pi = 0_\mathrm{SD1}^+$ and $0_\mathrm{SD2}^+$ bands.


\subsection{E2 transition strengths}

\begin{table}[tbp]
 \begin{center}
  \caption{
  Theoretical (left) and experimental (right) quadrupole electric transition strengths $B(\mathrm{E2})$ in Weisskopf units, $B_\mathrm{Wu}(\mathrm{E2}) = 6.54~\mathrm{e}^2\mathrm{fm}^4$, for positive-parity states.
  $J^\pi_\mathrm{i}$ and $J^\pi_\mathrm{f}$ indicate spin parity of initial and final states, respectively.
  $E_\mathrm{i}$ and $E_\mathrm{f}$ are excitation energies of initial and final states, respectively, in MeV.
  Experimental data are taken from Refs.~\onlinecite{PhysRevC.71.014316} and \onlinecite{Nica20121563}.
  }
 \label{BE2_positive}
  \begin{tabular}{cc}
   \begin{tabular}[t]{ccc}
    \multicolumn{3}{l}{Theory}\\
    \hline
    $J^\pi_\mathrm{i}$ & $J^\pi_\mathrm{f}$ & $B(\mathrm{E2})$\\
    \hline
    $2_\mathrm{1}^+$   & $0_\mathrm{1}^+$   & 11.03\\
    $2_\mathrm{2}^+$   & $0_\mathrm{1}^+$   & $<0.1$\\
    $2_\mathrm{2}^+$   & $2_\mathrm{1}^+$   & 5.90\\
    $3_\mathrm{1}^+$   & $2_\mathrm{1}^+$   & 0.16\\
    $3_\mathrm{1}^+$   & $2_\mathrm{2}^+$   & 5.99\\
    $4_\mathrm{1}^+$   & $2_\mathrm{1}^+$   & 8.39\\
    $4_\mathrm{1}^+$   & $2_\mathrm{2}^+$   & 0.20\\
    $4_\mathrm{1}^+$   & $3_\mathrm{1}^+$   & 3.30\\
    $2_\mathrm{ND}^+$  & $0_\mathrm{ND}^+$  & 18.39\\
    $4_\mathrm{ND}^+$  & $2_\mathrm{ND}^+$  & 16.09\\
    $6_\mathrm{ND}^+$  & $4_\mathrm{ND}^+$  & 19.75\\
    $8_\mathrm{ND}^+$  & $6_\mathrm{ND}^+$  & 31.06\\
    $2_\mathrm{SD1}^+$ & $0_\mathrm{SD1}^+$ & 110.63\\
    $4_\mathrm{SD1}^+$ & $2_\mathrm{SD1}^+$ & 156.24\\
    $6_\mathrm{SD1}^+$ & $4_\mathrm{SD1}^+$ & 168.96\\
    $2_\mathrm{SD2}^+$ & $0_\mathrm{SD2}^+$ & 152.02\\
    $4_\mathrm{SD2}^+$ & $2_\mathrm{SD2}^+$ & 217.08\\
    $6_\mathrm{SD2}^+$ & $4_\mathrm{SD2}^+$ & 238.99\\
    \hline
   \end{tabular} &
   \begin{tabular}[t]{ccccc}
    \multicolumn{5}{l}{Experiments}\\
    \hline
    $J^\pi_\mathrm{i}$ & $E_\mathrm{i}$ & $J^\pi_\mathrm{f}$ & $E_\mathrm{f}$ & $B(\mathrm{E2})$\\
    \hline
    $2^+ $ & 2.12  & $0^+$ & 0.00  & $6.24 \pm 0.16$ \\
    $2^+ $ & 3.30  & $0^+$ & 0.00  & $0.75 \pm 0.04$ \\
    $2^+ $ & 3.30  & $2^+$ & 2.12  & $3.8  \pm 1.0$ \\
    $2^+ $ & 4.11  & $0^+$ & 0.00  & $0.57 \pm 0.09$ \\
    $2^+ $ & 4.11  & $2^+$ & 2.12  & $2.3  \pm 0.6 $ \\
    $4^+ $ & 4.68  & $2^+$ & 2.12  & $8.2  \pm 1.4$ \\
    $3^+ $ & 4.87  & $2^+$ & 2.12  & $0.09 \pm 0.06$ \\
    $3^+ $ & 4.87  & $2^+$ & 3.30  & $0.8  \pm 0.8$ \\
    $2^+ $ & 4.88  & $0^+$ & 0.00  & $0.35 \pm 0.13$ \\
    $8^+ $ & 10.06 & $6^+$ & 8.50  & $27 \pm 15$ \\
    $9^+ $ & 12.14 & $7^+$ & 9.91  & $7.6 \pm 2.0$ \\
    $10^+$ & 13.34 & $8^+$ & 11.37 & $7.1 \pm 2.3$ \\
    \hline
   \end{tabular}\\
   \end{tabular}
 \end{center}
\end{table}

\begin{table}[tbp]
\begin{center}
 \caption{
 Same as Table~\ref{BE2_positive} but for negative-parity states.
 }
 \label{BE2_negative}
 \begin{tabular}{cc}
   \begin{tabular}[t]{ccc}
    \multicolumn{3}{l}{Theory}\\
    \hline
    $J^\pi_\mathrm{i}$ & $J^\pi_\mathrm{f}$ & $B(\mathrm{E2})$\\
    \hline
    $5_\mathrm{1}^-$   & $3_\mathrm{1}^-$   & 12.96\\
    $4_\mathrm{1}^-$   & $3_\mathrm{1}^-$   & 3.27\\
    $4_\mathrm{1}^-$   & $5_\mathrm{1}^-$   & 0.75\\
    $7_\mathrm{1}^-$   & $5_\mathrm{1}^-$   & 8.80\\
    $6_\mathrm{1}^-$   & $5_\mathrm{1}^-$   & 2.88\\
    $6_\mathrm{1}^-$   & $4_\mathrm{1}^-$   & 9.98\\
    $6_\mathrm{1}^-$   & $7_\mathrm{1}^-$   & 0.27\\
    $3_\mathrm{ND2}^-$ & $2_\mathrm{ND2}^-$ & 49.35\\
    $4_\mathrm{ND2}^-$ & $2_\mathrm{ND2}^-$ & 16.33\\
    $4_\mathrm{ND2}^-$ & $3_\mathrm{ND2}^-$ & 78.98\\
    $6_\mathrm{ND2}^-$ & $4_\mathrm{ND2}^-$ & 44.90\\
    $5_\mathrm{ND3}^-$ & $3_\mathrm{ND3}^-$ & 33.42\\
    $5_\mathrm{ND3}^-$ & $4_\mathrm{ND3}^-$ & 66.56\\
    $7_\mathrm{ND3}^-$ & $5_\mathrm{ND3}^-$ & 55.80\\
    $9_\mathrm{ND3}^-$ & $7_\mathrm{ND3}^-$ & 61.82\\
    $5_\mathrm{SD}^-$  & $4_\mathrm{SD}^-$  & 175.51\\
    $6_\mathrm{SD}^-$  & $4_\mathrm{SD}^-$  & 38.21\\
    $6_\mathrm{SD}^-$  & $5_\mathrm{SD}^-$  & 181.14\\
    $8_\mathrm{SD}^-$  & $6_\mathrm{SD}^-$  & 100.26\\
    \hline
   \end{tabular} &
   \begin{tabular}[t]{ccccc}
    \multicolumn{5}{l}{Experiments}\\
    \hline
    $J^\pi_\mathrm{i}$ & $E_\mathrm{i}$ & $J^\pi_\mathrm{f}$ & $E_\mathrm{f}$ & $B(\mathrm{E2})$\\
    \hline
    $5^- $ & 5.69  & $3^-$ & 4.62  & $0.76 \pm 0.12$ \\
    $6^- $ & 7.79  & $5^-$ & 5.69  & $14 \pm 4$ \\
    $6^- $ & 7.79  & $4^-$ & 6.25  & $16 \pm 6$ \\
    $7^- $ & 8.37  & $5^-$ & 5.69  & $7.4 \pm 1.6$ \\
    \hline
   \end{tabular}\\
 \end{tabular} 
\end{center}
\end{table}

Tables~\ref{BE2_positive} and \ref{BE2_negative} show the $B(\mathrm{E2})$ values for the positive- and negative-parity states, respectively.
The experimental values are also shown.
The in-band $B(\mathrm{E2})$ values of the $K^\pi = 0_\mathrm{SD1}^+$, $0_\mathrm{SD2}^+$, and $4_\mathrm{SD}^-$ bands are much larger than those of the other transitions, and the $B(\mathrm{E2})$ values are more than 100 $B_\mathrm{Wu}(\mathrm{E2})$, which indicates a large deformation of the $K^\pi = 0_\mathrm{SD1}^+$, $0_\mathrm{SD2}^+$, and $4_\mathrm{SD}^-$ bands.
The theoretical $B(\mathrm{E2}; 8_\mathrm{ND}^+\rightarrow 6_\mathrm{ND}^+)$ value is consistent with the experimental $B(\mathrm{E2}; 8^+ (10.06~\mathrm{MeV})\rightarrow 6^+ (8.50~\mathrm{MeV}))$ value.
\section{Discussion}
\label{discussions}

\subsection{Similarity of molecular orbitals around $^{16}$O + $^{16}$O cores and Nilsson orbits}

\begin{figure}[tbp]
 \begin{center}
  \includegraphics[width=0.5\textwidth]{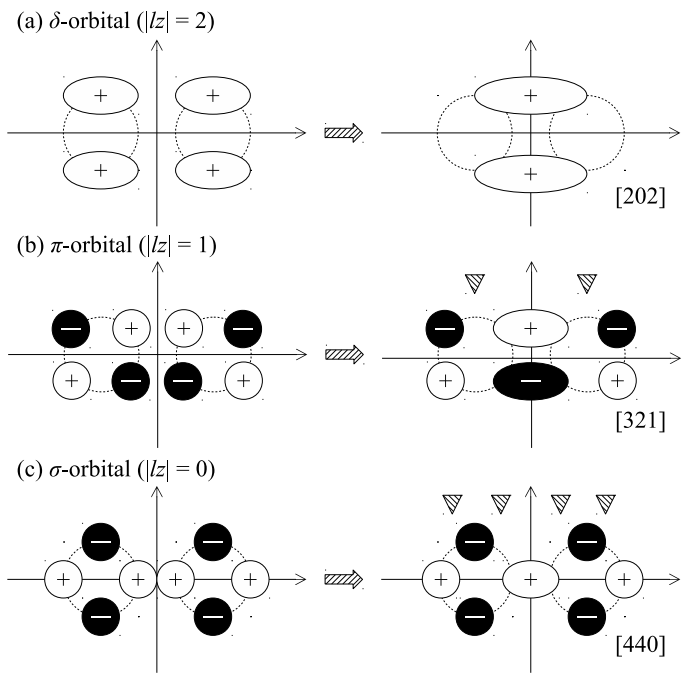}
  \caption{
  Schematic illustrations of molecular orbitals generated from $0d$ orbits around two $^{16}$O cores for (a) $\delta$, (b) $\pi$, and (c) $\sigma$ orbitals.
  Left and right columns show phases of $0d$ orbits around two $^{16}$O cores and molecular orbitals, respectively.
  Dotted circles show two $^{16}$O cores located on the $z$ axis.
  Inverse triangles show locations of nodes in molecular orbitals in the $z$ direction.
  Numbers in brackets show Nilsson quanta (see text).
  }
  \label{nilsson_molecular}
 \end{center}
\end{figure}

Before discussing the structures of the SD states in $^{34}$S, the quanta of molecular orbitals
around the $^{16}$O + $^{16}$O cores are discussed in the Nilsson picture.

The lowest orbit around an $^{16}$O core is a $0d_{5/2}$ orbit.
By linear combination of $0d_{5/2}$ orbits around two $^{16}$O cores, molecular
orbitals around the two $^{16}$O cores are formed.
Figure~\ref{nilsson_molecular} shows schematic illustrations of the formation of molecular orbitals
around two spherical $^{16}$O cores.
Here, the two $^{16}$O cores are located on the $z$ axis (horizontal axis), and a cylindrical coordinate system is used in this section because of axial symmetry around the $z$ axis.

The $0d_{5/2}$ orbits around the $^{16}$O cores form $\delta$, $\pi$, and $\sigma$ orbitals, which are formed from two $(l, |l_z|) = (2, 2)$, $(2, 1)$, and $(2, 0)$ orbits, respectively, around the left and right $^{16}$O cores.
For the $\delta$ orbital, the $(l, |l_z|) = (2, 2)$ orbits have no node in the $z$ direction, as shown in the left column of Fig.~\ref{nilsson_molecular}(a).
Therefore, the $\delta$ orbital also has no node in the $z$ direction, as shown in the right column of Fig.~\ref{nilsson_molecular}(a).
Similarly, the numbers of nodes of the $\pi$ and $\sigma$ orbitals are two and four in the direction of the $z$ axis, as shown in Figs.~\ref{nilsson_molecular}(b) and (c), respectively.
In the radial direction, they have no node because a $0d$ orbit has no node in the radial direction.
This shows that the quanta of the $\delta$, $\pi$, and $\sigma$ orbitals are [202], [321], and [440], respectively, in the Nilsson picture.
When a system has a two-$^{16}$O core structure, the [202], [321], and [440] orbits correspond to the $\delta$, $\pi$, and $\sigma$orbitals, respectively.
The parities of the $\delta$, $\pi$, and $\sigma$ orbitals are positive, negative, and positive, respectively.

\subsection{Configurations of valence neutrons in the SD bands}

The GCM calculation yielded three SD bands, called the $K^\pi = 0_\mathrm{SD1}^+$, $0_\mathrm{SD2}^+$, and $4_\mathrm{SD}^-$ bands, the structures of which are interpreted as $^{16}$O + $^{16}$O + two valence neutrons that have $\delta^2$, $\pi^2$, and $\delta^1 \pi^1$ configurations, respectively, in a cluster picture.

Their proton components have $2 \hbar \omega$ excited configurations and neck structures, and the neutrons, except for the two highest-energy orbits, have the same configuration.
This configuration is the same as that of the SD band in $^{32}$S, which contains many $^{16}$O + $^{16}$O cluster structure components\cite{PhysRevC.69.051304}.
Therefore, the three SD bands have structures of $^{16}$O + $^{16}$O + valence neutrons in a cluster picture.

The configurations of the valence neutrons of the $K^\pi = 0_\mathrm{SD1}^+$, $0_\mathrm{SD2}^+$, and $4_\mathrm{SD}^-$ bands are $[202]^2$, $[321]^2$, and $[202]^1[321]^2$, respectively.
As shown in the previous section, the $[202]$ and $[321]$ orbits correspond to the $\delta$ and $\pi$ molecular orbitals, respectively, around the two $^{16}$O cores.
Therefore, the configurations of the valence neutrons of the $K^\pi = 0_\mathrm{SD1}^+$, $0_\mathrm{SD2}^+$, and $4_\mathrm{SD}^-$ bands are interpreted as $\delta^2$, $\pi^2$, and $\delta^1\pi^1$, respectively.

The degeneracy of the [202] and [321] Nilsson orbits of neutrons in the highly deformed region (Figs.~\ref{spo_positive} and \ref{spo_negative}) and the coexistence of two positive- and one negative-parity SD bands (Fig.~\ref{level_scheme}) are explained in the cluster picture.
In this picture, the [202] and [321] orbits correspond to the $\delta$ and $\pi$ orbitals, 
which are formed by linear combination of $0d$ orbits around the two $^{16}$O cores.
When the $^{16}$O + $^{16}$O clustering develops, the $\delta$ and $\pi$ orbitals have similar energies.
The $^{16}$O + $^{16}$O cluster components in the $^{32}$S (SD) component generate similar energies for the [202] and [321] orbits in the highly deformed states of $^{34}$S.
Because of the coexistence of the [202] and [321] orbits, the $K^\pi = 0_\mathrm{SD1}^+$, $4_\mathrm{SD}^-$, and $0_\mathrm{SD2}^+$ bands coexist; they have $^{32}$S (SD) + $[202]^2_\nu$, $[202]^1_\nu[321]^1_\nu$, and $[321]^2_\nu$ configurations, respectively.


\subsection{Structures of positive- and negative-parity yrast states}

The structures of the yrast states vary dramatically.
The $K^\pi = 0_\mathrm{ND}^+$ and $0_\mathrm{SD1}^+$ band members appear as positive-parity yrast states, and the $K^\pi = 3_\mathrm{ND}^-$ band members appear as negative-parity yrast states.
By $\gamma$ spectroscopy experiments on high-spin states, it may be possible to observe the $K^\pi = 0_\mathrm{ND}^+$, $0_\mathrm{SD1}^+$, and $3_\mathrm{ND}^-$ bands.

In positive-parity states, the yrast states for $J^\pi \leq 4^+$ and $6 \leq J^\pi \leq 10$ have $0 \hbar \omega$ and $2 \hbar \omega$ configurations, respectively.
The $2\hbar\omega$ configuration of the yrast $J^\pi = 6^+$ and $8^+$ states are consistent with a shell model calculation\cite{PhysRevC.71.014316}.
The present calculation suggests that the yrast $J^\pi = 6^+$ (8.50 MeV) and $8^+$ (10.65 MeV) states are members of the $K^\pi = 0_\mathrm{ND}^+$ band.
The candidates for low-spin states in the $K^\pi = 0_\mathrm{ND}^+$ band are the $J^\pi = 0_2^+$ (3.92 MeV) and $2^+$ (4.89 MeV) states.
E2 transitions between them have been observed\cite{Moss1970577}, and the upper limit is $\simeq20$ $B_\mathrm{Wu}(\mathrm{E2})$, which is close to the theoretical $B(\mathrm{E2}; 2_\mathrm{ND}^+ \rightarrow 0_\mathrm{ND}^+)$ value.
To observe the low-spin members of the $K^\pi = 0_\mathrm{ND}^+$ band, in-band transitions from the $J^\pi = 6_1^+$ state are necessary.
In previous $\gamma$ spectroscopy experiments, the final state of the observed E2 transitions from the $J^\pi = 6_1^+$ state is only the $J^\pi = 4_1^+$ state, which is a $0\hbar \omega$ state.

In high-spin states, $K^\pi = 0_\mathrm{SD1}^+$ and $K^\pi = 3_\mathrm{ND}^-$ are yrast states of the positive- and negative-parity states for the $J^\pi \geq 12^+$ and $J^\pi \geq 9^-$ states, and the in-band $B(\mathrm{E2})$ values are large.
This shows that it may be possible to observe those bands by $\gamma$ spectroscopy experiments on high-spin states.
The observed states of the assigned spins and parities are limited to $J\leq 8$ states.
$\gamma$ spectroscopy experiments on the $J>12$ states are expected to reveal dramatic structural changes in $^{34}$S.

\section{Conclusions}
\label{conclusions}

The structure of the SD states in $^{34}$S were investigated using the AMD and GCM.
By superposing the AMD wave functions calculated via energy variation with a constraint on the quadrupole deformation parameter $\beta$, the coexistence of two positive- and one negative-parity SD bands is predicted.
The SD states have mp-mh configurations, and they are interpreted as a structure consisting of  $^{16}$O + $^{16}$O + valence neutrons in molecular orbitals around $^{16}$O + $^{16}$O cores in a cluster picture.
The structures of the yrast states vary dramatically.
The $K^\pi = 0_\mathrm{ND}^+$, $0_\mathrm{SD1}^+$, and $3_\mathrm{ND}^-$ band members appear as the yrast states of each parity.
Highly efficient $\gamma$ spectroscopy experiments on high-spin states may reveal the structures of those deformed states.

\begin{acknowledgments}
 The author thanks Dr. Ideguchi, Dr. Go, and Dr. Niikura for fruitful discussions. 
 This work was supported by JSPS KAKENHI Grant Number 25800124. 
 The numerical calculations for this work were conducted under the Interdisciplinary
Computational Science Program of the Center for Computational Sciences, University of Tsukuba.
\end{acknowledgments}

\bibliography{34S_v2}

\end{document}